\documentclass[aps,showpacs,prl,twocolumn]{revtex4}
\usepackage{amsmath,amssymb}
\usepackage{graphics}
\usepackage{graphicx}
\usepackage{color}

\begin{document}

\title{Statistics of Fourier Modes of Velocity and Vorticity in Turbulent Flows : Intermittency and Long-Range Correlations}

\author{L. Chevillard, N. Mazellier, C. Poulain, Y. Gagne and C. Baudet}
\affiliation{Laboratoire des Ecoulements G\'eophysiques et Industriels,\\
BP53, 38000 Grenoble, France}

\begin{abstract}
We perform a statistical analysis of experimental fully developed
turbulence longitudinal velocity data in the Fourier space. We
address the controversial issue of statistical intermittency of
spatial Fourier modes by acting on the finite spectral resolution.
We derive a link between velocity structure functions and the
flatness of Fourier modes thanks to cascade models. Similar
statistical behaviors are recovered in the analysis of spatial
Fourier modes of vorticity obtained in an acoustic scattering
experiment. We conclude that vorticity is long-range correlated
and found more intermittent than longitudinal velocity.

\end{abstract}

\pacs{02.50.Fz, 43.58.+z, 47.27.Gs}

\maketitle

In fully developed turbulence, most of the experimental, numerical
and theoretical works \cite{Fri95} focus on the statistics of the
longitudinal velocity increments $\delta_r u(x)=u(x+r)-u(x)$. It
is now well established that structure functions $M_q(r) = \langle
(\delta_r u)^q\rangle_x$, behave as power laws, i.e. $M_q(r)\sim
r^{\zeta_q}$. The universal non-linear evolution of $\zeta_q$ with
respect to $q$ is refered to the so-called intermittency
phenomenon: the probability density function (PDF) of velocity $u$
is close to Gaussian, while the PDF of longitudinal velocity
gradients $\partial_xu$ exhibits extremely large tails. Another
striking property of turbulence is the long-range correlation of
dissipation events, i.e. $(\partial_xu)^2$, up to the velocity
correlation length scale $L$. Many systems share same types of
behaviors, as financial volatility \cite{RefFinCas}(a) and
electrical transport in granular media \cite{RefFinCas}(b). A
major issue in turbulence is to derive a possible link between
long range correlations and vorticity filaments \cite{MoiJim04}.
In the Fourier space, which is an alternative way to study
turbulence statistics \cite{Pandit}, one could expect that the
Fourier modes of velocity $\widetilde{u}(k,t)$ should analogously
follow the same types of behaviors, i.e. $\langle
|\widetilde{u}(k,t)|^q\rangle_t \sim k^{-\zeta_q}$, since $k$ can
be interpreted as the inverse of a scale, as prescribed in shell
models \cite{Shell}. Furthermore, a statistical model based on the
rapid distortions of the small scales predicts strong
intermittency in the $k$-space \cite{DubJFM04}. Surprisingly, it
is not the case on experimental and numerical velocity profiles,
as pointed out by the seminal paper of Brun \& Pumir
\cite{BruPum01}, since the PDF of Fourier modes are found to be
undistinguishable from Gaussians, whatever the value of $k$.

However, one of the crucial parameters of the Fourier Transform is
the finite spectral resolution associated with the finite length
of velocity profiles hereafter noted $\ell$. The goal of this
letter is to show that, whereas statistics of Fourier modes do not
depend on $k$, they depends significantly on the ratio $\ell/L$.
Firstly, we present such a ``short time" Fourier analysis of
experimental longitudinal velocity data. Secondly, the statistical
dependence on $\ell/L$ is clarified in the context of various
turbulent cascade models. Finally, we perform a similar analysis
on experimental data, obtained in an acoustic scattering
experiment, allowing the direct probing in time of spatial Fourier
modes of vorticity in a turbulent air jet. By comparison with the
longitudinal velocity data analysis, we conclude that vorticity is
also long-range correlated but more intermittent.

\begin{figure}
\center{\includegraphics[width=8cm]{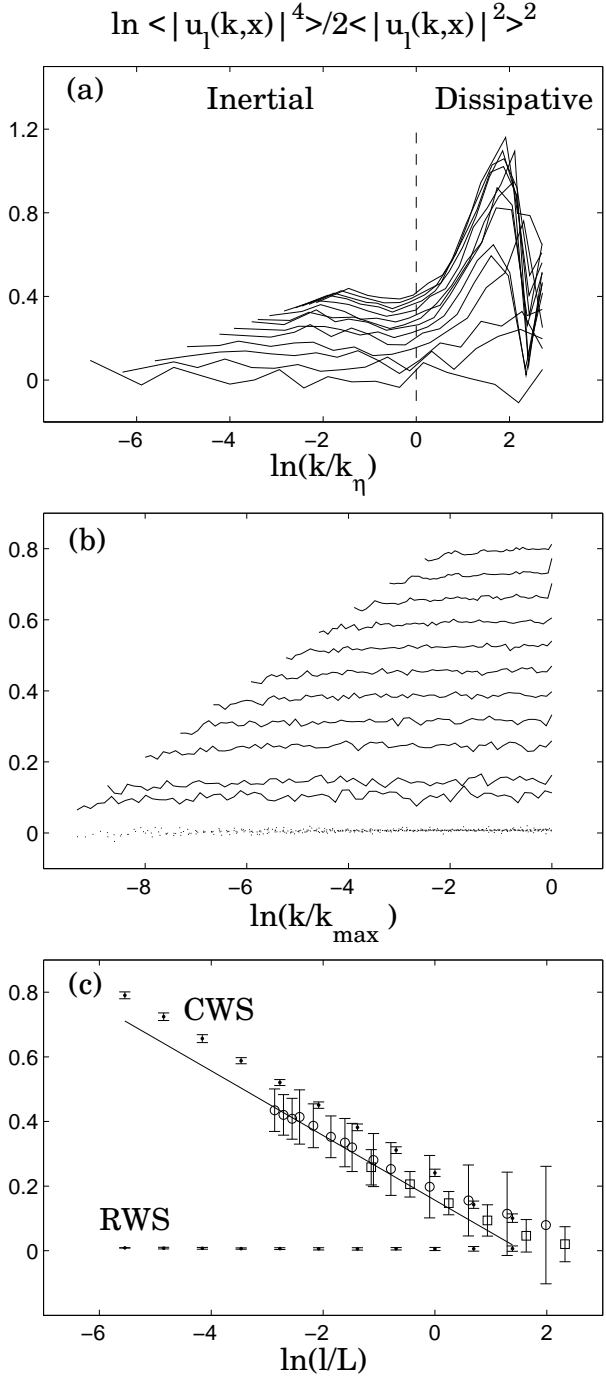}}
\caption{\label{fig:StatFlat} (a) Flatness of $|\widetilde{u}_\ell
(k,x)|$ as a function of $\ln (k/k_\eta)$
($k_\eta=(\eta_K)^{-1}$), for different window lengths $\ell$
(from top to bottom, $\ell/L =$ 0.057; 0.067; 0.078; 0.089; 0.114;
0.156; 0.200; 0.228; 0.334; 0.456; 0.91; 1.82; 3.64; 7.28) for the
Modane velocity signal ($\mathcal R_\lambda \approx 2500)$.(b)
Flatness of synthetic velocity profiles $|\widetilde{u}^{{\small
\mbox{rws}}}_\ell (k,x)|$ (dotted line) and
$|\widetilde{u}^{{\small \mbox{cws}}}_\ell (k,x)|$ (solid line)
vs. $\ln (k/k_{{\small \mbox{max}}})$, where $k_{{\small
\mbox{max}}}$ is the Nyquist wave vector and for several scales
(from top to bottom $\log_2(\ell_p/L) =$ -8, -7, -6, -5, -4, -3,
-2, -1, 0, 1 and 2). (c) Values of the plateaus of Figs. (a) and
(b) vs. $\ln (\ell/L)$ : ($\circ$) Modane velocity data,
($\square$) air-jet velocity data ($\ell/L=$ 0.32; 0.64; 1.28;
2.56; 5.12; 10.24), ($\bullet$) synthetic velocities, (solid line)
our theoretical prediction (Eq. (\ref{eq:FlatCWS})).}
\end{figure}

Let us introduce the Short-Time Fourier analysis
$\widetilde{u}_\ell (k,x)$, which depends on the space variable
$x$ and the wave vector $k$ and corresponds to the computation of
the Fourier transform of the longitudinal velocity $u(x)$ in a
window $h_\ell$ of size $\ell$, i.e. $\widetilde{u}_\ell (k,x) =
u(x)\otimes \left(e^{-ikx}h_\ell(x)\right)$, where $\otimes$
stands for the convolution product. We represent in Fig.
\ref{fig:StatFlat}(a) the flatness of $|\widetilde{u}_\ell
(k,x)|$, i.e. $\mathcal F_\ell(k) = \langle |\widetilde{u}_\ell
(k,x)|^4\rangle /\langle |\widetilde{u}_\ell (k,x)|^2\rangle ^2 $,
as a function of $\ln (k/k_\eta)$, where $k_\eta=(\eta_K)^{-1}$
($\eta_K$ is the dissipative Kolmogorov length scale), for several
windowing length $\ell$ (see the caption), and for
$k\ge2\pi/\ell$. The experimental longitudinal velocity signal
used in this study has been recorded in the Modane's wind tunnel
facility \cite{KahMal98} at a Taylor microscale Reynolds number
$\mathcal R_\lambda \approx 2500$ and thus exhibits a large
inertial range. Here, the spectral window is the Hanning function:
$h_\ell(x)=\cos ^2(\pi x/\ell)$ for $x\in [-\ell/2,\ell/2]$,
$h_\ell(x)=0$ instead. For inertial wave vectors ($k<k_\eta$),
$\mathcal F_\ell(k)$ slightly depends on $k$ but drastically on
$\ell$: when $\ell$ is of order of several correlation lengths
($\ell \gg L$), $\mathcal F_\ell(k)$ is close to the Gaussian
value 2 (as found in \cite{BruPum01}) and when $\ell/L\rightarrow
0$, we note a rapid increase of the value of this inertial
``plateau". The evolution of $\mathcal F_\ell(k)$ with respect to
$\ell/L$ is displayed on Fig. \ref{fig:StatFlat}(c). The error
bars have been obtained by a least-square fit of the plateaus of
the flatness over wave vectors in the inertial range. We thus
observe that the flatness of the Fourier modes behaves as a power
law of the scale $\ell$, i.e. $\mathcal F_\ell(k) \sim \ell
^{\alpha}$ with $\alpha= -0.1 \pm 0.02$. Here, the error bar
$0.02$ is large because of the lack of statistics and the fact
that longitudinal velocity profiles, obtained under the Taylor
Hypothesis \cite{Fri95}, are sensitive to the temporal
decorrelation \cite{Cas02}. We have also displayed on Fig.
\ref{fig:StatFlat}(c) the evolution of the flatness for an another
longitudinal velocity profile ($R_\lambda \approx 300$) obtained
in the air jet we will present at the end of this letter. Hence,
$\alpha$ remains unchanged and can be considered as universal. Let
us mention that the second moment of $|\widetilde{u}_\ell (k,x)|$
is obviously proportional to $k^{-5/3}$ and $\langle
|\widetilde{u}_\ell (k,x)|^2 \rangle = (\ell/L)\langle
|\widetilde{u}_L (k,x)|^2 \rangle$ for $k\ge 2\pi/\ell$. In the
dissipative range (i.e. $k>k_\eta$), the flatness appears to
rapidly increase without any saturation \cite{CheCas03}. This is
one of the first experimental verification of a Kraichnan's
conjecture \cite{kraichnan,FriMor81} which is linked to the
log-infinitely distribution breaking of velocity \cite{saito92}
and is currently under investigations. In the following, we will
theoretically establish a link between the inertial exponent
$\alpha$ and the structure function exponent $\zeta_q$ :
$\alpha=\zeta_4-2\zeta_2$.

Let us now begin with defining an intermittent (zero mean)
velocity profile. This is usually done with the help of wavelet
series \cite{OrthoWavMath}, early introduced in the context of
turbulence \cite{Men91,Benetal93},
\begin{equation}\label{eq:WavSer}
u(x) = \sum_{j=0}^{+\infty}\sum_{k=0}^{2^j-1}c_{j,k}\psi_{j,k}(x)
\mbox{ ,}
\end{equation}
where the set $\{\psi_{j,k}(x) = 2^{j/2}\psi(2^jx-k) \}$ is an
orthonormal basis of the space of finite energy functions
$L^2([0,L])$ (see \cite{ArnWav}) and $\psi$ an admissible
``mother" wavelet. The wavelet coefficients $c_{j,k}$ govern the
statistics across scales. Generally, the coefficients $c_{j,k} =
2^{-j/2}\epsilon_{j,k}\beta_{j,k}$ are chosen as a product of a
sign ($\epsilon_{j,k} = \pm 1 $ with equal probability) and
positive random variables $\beta_{j,k}$ that are chosen so as to
be compatible with turbulence longitudinal velocity statistics,
i.e. $\mathbb E (\beta_{j,k}^q)= 2^{-j\zeta_q}$ ($\mathbb E(.)$
meaning here mathematical expectation). Moreover, as already
predicted by the unifying point of view of Cates \& Deutsch
\cite{CatDeu87}, statistics of velocity fluctuations are
correlated in space and scale, that can be formalized through
space-scale correlations of dyadic wavelet coefficients as
$\mathbb E (\beta_{j,k} ^{q_1} \beta_{l,m}^{q_2}) = \mathbb E
(\beta_{j,k} ^{q_1})\mathbb E(\beta_{l,m}^{q_2}) \mathcal
C^{q_1,q_2}_{j,k,l,m}$ where the functions $\mathcal
C^{q_1,q_2}_{j,k,l,m}$ render additional correlations
\begin{equation}\label{eq:LongRangCorr}
\mathcal C^{q_1,q_2}_{j,k,l,m}=\left[ |k2^{-j}-m2^{-l}| + \max
(2^{-j},2^{-l})\right]^{\zeta_{q_1+q_2}-\zeta_{q_1}-\zeta_{q_2}}
\end{equation}
stating that wavelet coefficients are typically correlated, in
amplitude, up to the correlation length $L$. The generated
velocity profile $u^{{\small \mbox{cws}}}(x)$ using Eq.
(\ref{eq:WavSer}), where $\beta_{j,k}$ are correlated according to
Eq. (\ref{eq:LongRangCorr}) will be called a \textit{Cascade
Wavelet Series} (CWS) \cite{BruPum01,Benetal93,ArnWav}. Using the
simplest admissible Haar wavelet (i.e. $\psi(x)=1$ for $x\in
[0;L/2[$, $\psi(x)=-1$ for $x\in [L/2;L[$, $\psi(x)=0$ instead)
and the box for the Short Time Fourier transform (i.e. $h_\ell(x)
= 1$ for $x\in [0;\ell]$, and $h_\ell(x) = 0$ instead), it can be
shown analytically that $\mathbb E (|\widetilde{u}^{{\small
\mbox{cws}}}_{\ell_p} (k_n,0)|^2 ) \propto
(\ell_p/L)k_n^{-1-\zeta_2}$, for $k_n=\pi 2^{n}/L>\pi/\ell_p$ and
$\ell_p = L2^{-p}$ ($(n,p)\in \mathbb N ^2$). Moreover the
flatness $\mathcal F_{\ell_p}(k_n)=\mathbb E
(|\widetilde{u}^{{\small \mbox{cws}}}_{\ell_p}
(k_n,0)|^4)/(\mathbb E (|\widetilde{u}^{{\small
\mbox{cws}}}_{\ell_p} (k_n,0)|^2))^2$ behaves as
\begin{equation}\label{eq:FlatCWS}
\mathcal F_{\ell_p}(k_n) = 2
\frac{2}{(1+\zeta_4-2\zeta_2)(2+\zeta_4-2\zeta_2)} \left(
\frac{\ell_p}{L}\right)^{\zeta_4-2\zeta_2}
\end{equation}
when ${\ell_{p}\rightarrow 0}$ independently of $k_n$. We thus
have demonstrated that $\alpha=\zeta_4-2\zeta_2$. Note that the
flatness is not exactly equal to 2 at the correlation length
($\ell_p=L$).

In order to check our analytical predictions, in particular to
verify whether our computations depend on the box $h_\ell$ and the
synthesis wavelet $\psi$, we perform a statistical study of the
process $u^{{\small \mbox{cws}}}(x)$, using a more regular
Daubechies-6 wavelet for the synthesis wavelet $\psi$ and a
Hanning window for $h_\ell$. The method used to build the positive
random variables $\beta_{j,k}$ is the classical multiplicative
cascade model
\cite{BruPum01,Benetal93,ArnWav,CatDeu87,MenSre87,OllPar98} :
recursively, $\beta_{0,0} = 1$,
$\beta_{j,2k}=W_{j-1,k}^{(l)}\beta_{j-1,k}$ and
$\beta_{j,2k+1}=W_{j-1,k}^{(r)}\beta_{j-1,k}$, where the
$W_{j-1,k}^{(\kappa)}$ ($\kappa$ = $l$ for \textit{left} or $r$
for \textit{right}) are independent identically distributed
(i.i.d) positive random variables (see \cite{Benetal93,ArnWav}).
As an example, we will study the log-normal case where each $\ln
W_{j-1,k}^{(\kappa)}$ is a Gaussian random variable of mean
$\mu\ln 2$ and variance $\sigma^2\ln 2$ (leading to the quadratic
spectrum $\zeta_q = \mu q -\sigma^2 q^2/2$). We have used the set
of parameters $\sigma^2=0.025$ and $\mu = 1/3+3\sigma^2/2$,
consistent with experimental findings \cite{ArnetAl}, so that
$\zeta_2 \approx 2/3$, $\zeta_3=1$ and $2\zeta_2-\zeta_4=0.1$.
Numerically, the infinite sum in Eq. (\ref{eq:WavSer}) is
truncated at $j=N=2^{18}$ and is generated over $2^5$ integral
scales. It can be shown that such a stochastic process is not
stationary \cite{MenCha90} but at first order, its correlation
function is consistent with Eq. (\ref{eq:LongRangCorr}). We show
on Fig. \ref{fig:StatFlat}(b) (solid line) the estimation of the
flatness of $\widetilde{u}^{{\small \mbox{cws}}}_{\ell} (k,x)$ as
a function of $\ln (k/k_{{\small \mbox{max}}})$. After a
k-dependent crossover (data not shown) linked, among other
reasons, to the effect of discretness in the cascade
\cite{OllPar98}, the flatness does not depend on the wave vector
$k$ but significantly depends on $\ell_p$. In Fig.
\ref{fig:StatFlat}(c), we have gathered all the values of the
inertial plateaus using a least-square fit providing an error bar
estimation. The plateau behaves as a power law of the scale
$\ell_p$ in accordance with Eq. (\ref{eq:FlatCWS}) :
$\zeta_4-2\zeta_2 = -0.1 \pm 0.001$. The discrepancies between the
prefactors are mainly linked to the non-stationary character of
this generated synthetic velocity profile.

We would like to mention that if wavelet coefficients are no
longer long-range correlated (take $\mathcal
C^{q_1,q_2}_{j,k,l,m}=1$) and if $\ln \beta_{j,k}$ are chosen to
be independent Gaussian random variables with mean $\mu \ln 2^j$
and variance $\sigma ^2 \ln 2^j$, the corresponding synthetic
velocity generated will be called a \textit{Random Wavelet Series}
${u}^{{\small \mbox{rws}}}(x)$  which is intermittent in a
mathematical sense \cite{AubJaf02}. By construction, $\mathbb E
(|\widetilde{u}^{{\small \mbox{rws}}}_{\ell_p} (k_n,0)|^2
)=\mathbb E (|\widetilde{u}^{{\small \mbox{cws}}}_{\ell_p}
(k_n,0)|^2 )$, and analytical calculations performed in the same
framework defined in the context of CWS (i.e. using a Haar wavelet
for $\psi$ and a box for $h_\ell$) shows that the flatness
$\mathcal F_p(k_n)$ is equal to $2$ for both every wave vectors
$k_n$ and window lengths $\ell_p$. This property has been checked
numerically (with the same parameters $\mu$ and $\sigma$, and the
same synthesis wavelet $\psi$ and analysis window $h_\ell$ as in
the CWS case). The results are presented in Figs.
\ref{fig:StatFlat}(b) and (c). This heuristic synthetic process
shows that experimental longitudinal velocity data are not only
intermittent, but also long-range correlated.
\begin{figure}[t]
\center{\includegraphics[width=8cm]{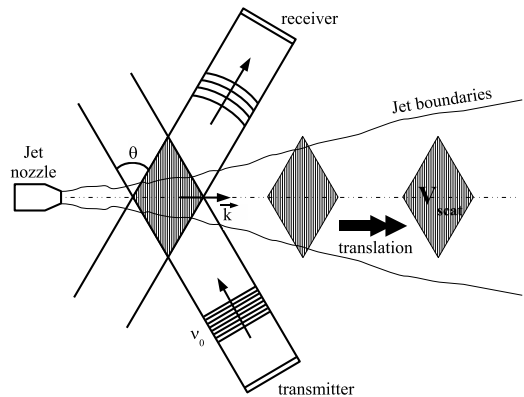}}
\caption{Acoustic scattering experiment in a turbulent jet flow. }
\label{fig:DispoExp}
\end{figure}

\begin{figure}[t]
\centerline{\includegraphics[width=8cm]{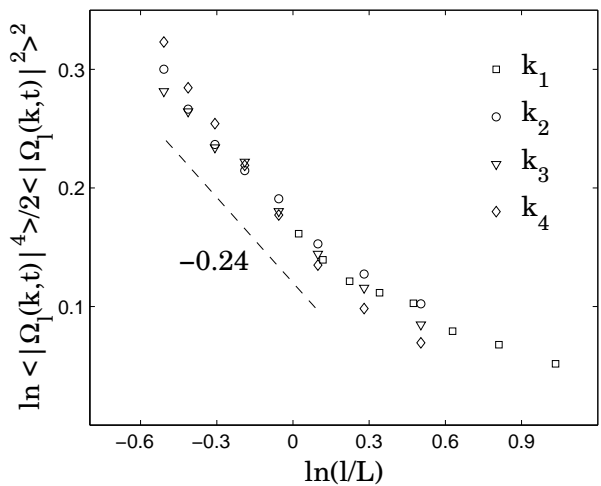}}
\caption{\label{fig:FlatExp} Flatness of experimental Fourier
modes of vorticity vs. $\ln (\ell/L)$, for four inertial wave
vectors $k_i$. Dotted line slope equal to -0.24. }
\end{figure}

We will turn now to acoustic scattering measurements allowing the
direct access to a spectral characterization of the vorticity
distribution. The non-linear interaction of an acoustic wave with
a turbulent flow gives rise to a scattering process of the
incident sound wave by the turbulent vorticity distribution. As in
any scattering experiment (like e.g. light or neutron scattering),
the complex amplitude of the scattered acoustic pressure field is
directly related to the incident acoustic amplitude and to the
spatial Fourier transform of the vorticity distribution
\cite{LunRoj89}. We have performed such a spectral investigation
on a turbulent axisymetric jet in air, at a Taylor based Reynolds
number $R_\lambda \approx 300$. We use a bistatic configuration
(Fig. \ref{fig:DispoExp}) wherein a plane monochromatic acoustic
wave with frequency $\nu_0$ and complex amplitude $p_{inc}(t)$ is
directed on the turbulent flow. The complex amplitude $
p_{scatt}(t)$ of the sound wave scattered at angle $\theta$ is
then recorded along time by a separate acoustic receiver. Further
details of the experimental apparatus can be found elsewhere
\cite{PouMaz04}. The scattering process results in phase and
amplitude modulations of the scattered acoustic pressure with
respect to the incident one: $ p_{scatt}(t)
=\widetilde{\Omega}_\perp\left(\vec{k},t\right).p_{inc}(t)$ where,
\begin{equation}\label{eq:ScattAmplitude}
\widetilde{\Omega}_\perp\left(\vec{k},t\right)  =
\iiint_{\mbox{V}_{{\small \mbox{scatt}}}}
\Omega_\perp\left(\vec{x},t\right) e^{-i \vec{k}\cdot\vec{x}} d^3x
\end{equation}
is the spatial Fourier transform of the vorticity component normal
to the scattering plane at wave-vector $\vec{k}$ such that
$|\vec{k}| = 4 \pi \nu_0/c\times \sin\left(\theta/2\right)$, with
$c$ the sound speed. By fixing both $\nu_0$ and $\theta$, the
scattering experiment allows the direct probing, in time, of a
well defined spatial Fourier mode of the turbulent vorticity
distribution characterized by a unique spatial wave-vector
$\vec{k}$ (spectral resolution). The price to pay for such a
spectral resolution lies in some spatial delocalization in the
physical space manifesting itself as a windowed spatial Fourier
transform over a finite volume ${\mbox{V}_{{\small
\mbox{scatt}}}}$ according to equation (\ref{eq:ScattAmplitude}).
The measurement volume is defined by the intersection of the
incident and detected acoustic beams and mainly depends on
$\theta$ and on the size of both acoustic transducers. In the
present experiment, $\theta = 60^{o}$ and the diameter of the
circular transducers is $14 cm$, leading to spatial resolutions of
order the integral length scale of the jet flow $L$. By varying
$\nu_0$, at a fixed $\theta$, four different wave-vectors $k_i$,
$i=1,2,3,4$ (in growing order) have been analysed, corresponding
to various length scales spanning the whole inertial range of the
turbulent flow. In the spirit of the first part of this letter, we
are interested in the influence of the parameter $\ell/L$ where
$\ell \sim \left(\mbox{V}_{{\small \mbox{scat}}}\right)^{1/3}$ is
a typical size of the measurement volume and $L$ is the integral
scale of the flow over the flatness $\mathcal F_\ell(k_i) =
\langle |\Omega_\ell(\vec{k},t)|^4\rangle_t/\langle
|\Omega_\ell(\vec{k},t)|^2\rangle_t ^2$. To this end, we rely on
the classical selfsimilarity property of the axisymetric turbulent
jet \cite{Wygnanski} according to which the integral scale $L$
(and all other pertinent scales) increases linearly with respect
to the distance downstream from the jet nozzle. A well known
consequence of this statistical selfsimilarity is the
invariability of the Reynolds number for large enough distances
from the jet nozzle. Several scattering experiments have been
performed at different distances from the jet nozzle,
corresponding to different integral length scales $L$. On Fig.
\ref{fig:FlatExp} is plotted the flatness of the modal amplitude
$\mathcal F_\ell(k_i)$ as a function of $\ln (\ell/L)$. Firstly,
$\mathcal F_\ell(k_i)$ does not depend on the wave vector $k_i$ at
a first order for $\ell<L$, in accordance with Fig.
\ref{fig:StatFlat}(c). Secondly, we see that the flatness behaves
as a power law with scale $\ell$, i.e. $\mathcal F_\ell(k_i)\sim
\ell ^{\gamma}$ with $\gamma = -0.24 \pm 0.02$ when $\ell/L
\rightarrow 0$.
Analogously with longitudinal velocity, the exponent $\gamma$ is
directly related to the classical $\zeta_q^{\Omega}$ exponent of
structure functions of vorticity considered as a vector field,
namely $\gamma=\zeta_4^{\Omega}-2\zeta_2^{\Omega}$. It is
noticeable that our experimental finding is in excellent agreement
with a tensorial wavelet analysis of Kestener \& Arneodo
\cite{KesArn04} that have been applied to a 3D-vorticity field
obtained from a Direct Numerical Simulation (DNS) of Navier-Stokes
equations at a smaller Reynolds number $\mathcal R_\lambda=140$
for which they obtained $\zeta_4^{\Omega} -2\zeta_2^{\Omega} =
-0.22 \pm 0.016$. In the same spirit, it has been measured that
tranverse velocity profiles are more intermittent that
longitudinal ones \cite{DhrTsu97}.

To sum up, we have shown that longitudinal velocity and vorticity
are intermittent and long-range correlated in the physical space
thanks to the study of the flatness of experimental velocity data
and acoustical measurements of vorticity Fourier modes. We have
seen that vorticity is much more intermittent than longitudinal
velocity. Our theoretical study on wavelet series shows that the
intermittency is intrinsically related to long-range correlations.
We mention the necessity to generalize this approach to
dissipative length-scales and wave vectors for which
log-infinitely divisible principles are violated.

This work is supported by the French Research Ministery, Joseph
Fourier University (PPF plateforme exp\'erimentale de
spectroscopie acoustique multi-\'echelles) and the CNRS. We wish
to acknowledge B. Lashermes and P. Borgnat for useful comments.



\begin{thebibliography}{99}

\bibitem{Fri95}
U. Frisch, \textit{Turbulence}, Cambridge University Press,
Cambridge (1995).
\bibitem{RefFinCas}
(a) I. Giardina \textit{et al.}, Physica A \textbf{299}, 28
(2001). (b) E. Falcon \textit{et al.}, Europhys. Lett.
\textbf{65}, 186 (2004).
\bibitem{MoiJim04}
F. Moisy and J. Jim\'enez, J. Fluid Mech. \textbf{513}, 111
(2004).
\bibitem{Pandit}
S. K. Dhar \textit{et al.}, Phys. Rev. Lett. \textbf{78}, 2964
(1997).
\bibitem{Shell}
L. Biferale, Ann. Rev. Fluid. Mech. \textbf{35}, 441
(2003).
\bibitem{DubJFM04}
B. Dubrulle \textit{et al.}, J. Fluid Mech. \textbf{520}, 1
(2004).
\bibitem{BruPum01}
C. Brun and A. Pumir, Phys. Rev. E \textbf{63}, 056313 (2001).
\bibitem{KahMal98}
H. Kahalerras \textit{et al.}, Phys. Fluids \textbf{10}, 910
(1998).
\bibitem{Cas02}
B. Castaing, Eur. Phys. J. B \textbf{29}, 357 (2002).
\bibitem{CheCas03}
L. Chevillard \textit{et al.}, Eur. Phys. J. B \textbf{45}, 561
(2005).
\bibitem{kraichnan}
R.~H. Kraichnan, Phys. of Fluids \textbf{10}, 2080 (1967).
\bibitem{FriMor81}
U. Frisch and R. Morf, Phys. Rev. A \textbf{23}, 2673 (1981).
\bibitem{saito92}
Y.~Saito, Phys. Soc. Japan \textbf{61}, 403 (1992). E.~A. Novikov,
Phys. Rev. E \textbf{50}, 3303 (1994).
\bibitem{OrthoWavMath}
Y. Meyer, \textit{Ondelettes} (Hermann, Paris, 1990). I.
Daubechies, \textit{Ten Lectures on Wavelets} (S.I.A.M.,
Philadelphia, 1992).
\bibitem{Men91}
C. Meneveau, J. Fluid Mech. \textbf{232}, 469 (1991).
\bibitem{Benetal93}
R. Benzi \textit{et al.}, Physica (Amsterdam) \textbf{65D}, 352
(1993).
\bibitem{ArnWav}
A. Arneodo \textit{et al.}, J. Math. Phys. \textbf{39}, 4142
(1998). A. Arneodo \textit{et al.}, Phys. Rev. Lett. \textbf{80},
708 (1998).
\bibitem{CatDeu87}
M. E. Cates and J. M. Deutsch, Phys. Rev. A \textbf{35}, 4907
(1987).
\bibitem{MenSre87}
C. Meneveau and K. R. Sreenivasan, Phys. Rev. Lett. \textbf{59},
1424 (1987).
\bibitem{OllPar98} P. Olla and P.
Paradisi, chao-dyn/9803039.
\bibitem{ArnetAl}
A. Arneodo \textit{et al.}, Europhys. Lett. \textbf{34}, 411
(1996).
\bibitem{MenCha90}
J. O'Neil and C. Meneveau, Phys. Fluids A \textbf{5}, 158 (1993).
\bibitem{AubJaf02} J.-M. Aubry and
S. Jaffard, Commun. Math. Phys. \textbf{227}, 483 (2002).
\bibitem{LunRoj89}
R.H. Kraichnan, J. Acoust. Soc. Am. \textbf{25}, 1096 (1953). F.
Lund and C. Rojas, Physica D \textbf{37}, 508 (1989).
\bibitem{PouMaz04} C. Poulain \textit{et
al.}, Flow, Turb. Comb. \textbf{72}, 245 (2004). B. Dernoncourt
\textit{et al.}, Physica D \textbf{117}, 181 (1998).
\bibitem{Wygnanski} I. Wygnanski and H. Fiedler, J. Fluid. Mech.
\textbf{38} (3), 577 (1969).
\bibitem{KesArn04}
P. Kestener and A. Arneodo, Phys. Rev. Lett. \textbf{93}, 044501
(2004).
\bibitem{DhrTsu97}
B. Dhruva \textit{et al.}, Phys. Rev. E \textbf{56}, R4928 (1997).




\end{thebibliography}
\end{document}